\documentclass[prb]{revtex4}

\usepackage{graphicx}
\usepackage{dcolumn}
\usepackage{bm}
\begin{document}

\title{Long Josephson junctions with spatially inhomogeneous driving}
\author{Andrey L. Pankratov}
\affiliation{Institute for Physics of Microstructures of RAS,
Nizhny Novgorod, RUSSIA. E-mail: alp@ipm.sci-nnov.ru}

\begin{abstract}
The phase dynamics of a long Josephson junction with
spatially inhomogeneously distributed bias current
is considered for the case of a dense soliton chain (regime of the
Flux Flow oscillator). To derive the analytical solution of the
corresponding sine-Gordon equation the Poincare method has been
used. In the range of the validity of the theory good coincidence
between analytically derived and numerically computed
current-voltage characteristics have been demonstrated for the
simplest example of unitstep function distribution of bias
current (unbiased tail). It is shown, that for the considered
example of bias current distribution, there is an optimal length
of unbiased tail that maximizes the amplitude of the main
harmonic and minimizes the dynamical resistance (thus leading
to reduction of a linewidth).

\noindent PACS number(s): $74.50.+r$
\end{abstract}

\maketitle

\section{Introduction}

Long Josephson oscillators operating in the flux-flow regime
\cite{nei} are presently considered as possible devices for
applications in superconducting millimeter-wave electronics
\cite{ksh},\cite{km}. In comparison to single fluxon oscillators
they have higher output power, wider bandwidth, and easier
tunability, but they have a wider linewidth of the emitted
radiation from the junction \cite{lwss},\cite{kosh},\cite{kosh2}.
The phase dynamics in long Josephson junctions has been
intensively studied both numerically and analytically. As
classified in \cite{cdlsg97}, the studies can be divided into two
categories. In the first category one can find sine-Gordon
solitons created by a small magnetic field confined to the
boundaries of the junction and propagating along the junction
under the influence of the bias current. In this case one can extract
detailed information about
the physics of the problem from the McLaughlin and Scott theory
\cite{scott} as the aim of that theory was to investigate the
steady-state motion of fluxons (and other solutions of the
sine-Gordon equation) under small perturbations. In the second
category one can find dense soliton chains distributed along one
dimension of the junctions driven by dc and (or) ac currents. The
physics of this category cannot be treated by the
McLaughlin-Scott perturbation theory for solitons as the
perturbations (bias current and magnetic field) are no longer
small compared to the sine-Gordon terms. However, namely this
regime is interesting for practical flux flow oscillators
\cite{ksh},\cite{km},\cite{kosh},\cite{kosh2}.
The latter regime is called the "flux-flow" regime and is
characterized by excitations which travel on top of a fast
rotating background so that the effective nonlinearity in the
system is drastically reduced due to fulfilling the following
conditions: $\eta_0/\alpha \gg 1$ and (or) $\Gamma\gg 1$ (where
$\eta_0$, $\alpha$ and $\Gamma$ are dimensionless total dc bias
current through the junction, the damping and the magnetic field,
respectively). It should be noted, that the "flux-flow" regime
has nothing to do with motion of well-distinguishable
flux-quanta: the soliton chain is so dense, that one should speak
about transmission of quasi-linear waves in a long Josephson
junction in this case.

For the flux-flow regime two main approaches are known. One is
the approach by Kulik \cite{kulik} (that has also been used, e.g., in
\cite{cgssv98}, \cite{ss99}), that is based on linear mode theory and
perturbative analysis around rotating background
($\phi=\phi_0+\psi$, $\psi\ll 1$). Another approach has been
suggested in Ref. \cite{jaw} and is also based on known form
of a solution around which one can make a perturbative expansion,
but in a different manner: the unidirectional fluxon train is
accompanied by two plasma waves, that allows to satisfy boundary
conditions. Using either of the approaches one can compute
the current-voltage characteristic of FFO in the second order
approximation for spatially homogeneous dc and ac driving.
However, if the bias current has inhomogeneous spatial
distribution, it is not clear how to derive the corresponding
solutions, since the approaches are based on the "anzats", some
known form of the initial solution, around which the perturbative
analysis has to be done. It is known from
\cite{Zhang} that the use of inhomogeneous driving
(unbiased tail) may lead to decrease of a dynamical
resistance and therefore to reduction of a linewidth.
Also recently importance of accounting of
magnetic field fluctuations for the analysis and design of FFO has
been experimentally demonstrated \cite{koshisec}. Further
theoretical \cite{lwp} and experimental \cite{koshpc}
investigations have shown that not only usual dynamical resistance
$R_d=dV/dI$, but also the dynamical resistance with respect to
the magnetic field $R_d^{H}=dV/dH$ (that may also be
attributed to the dynamical resistance of control line
$R_d^{CL}=dV/dI_{CL}$) is important for calculation of the
linewidth of FFO. Therefore in FFO designs one should care about
minimization of both above mentioned dynamical resistances
and the present theoretical state of the problem should
be reconsidered in order to improve characteristics
of practical flux flow oscillators since nobody payed attention
before to the value of $R_d^{H}$.

The aim of this paper is to present the approach that allows
to systematically study the regime of large magnetic fields and
bias currents and to describe the dynamics of the Flux Flow
oscillator with spatially inhomogeneous driving (and, as
particular example, with unbiased tail). This approach allows to
perform the detailed analysis of FFO and to find an
optimal profile of the bias current distribution and an optimal
length of unbiased tail in order to maximize the output power of
FFO and to reduce the linewidth.

\section{Basic equations and the Poincare method}

The electrodynamics of a long Josephson junction in the presence
of magnetic field is described by the perturbed sine-Gordon
equation
\begin{equation}
\frac{\partial^2\phi}{\partial t^2}+\alpha\frac{\partial\phi}
{\partial t}-\frac{\partial^2\phi}{\partial x^2}=\eta(x)-\sin (\phi)
\label{PSGE}
\end{equation}
subject to the boundary conditions
\begin{equation}
\frac{\partial\phi(0,t)}{\partial x}=
\frac{\partial\phi(L,t)}{\partial x}=\Gamma.  \label{bc}
\end{equation}
In this equation space and time have been normalized to the
Josephson penetration length $\lambda _{J}$ and to the inverse
plasma frequency $\omega_{p}^{-1}$, respectively, $\alpha$ is the
loss parameter, $\eta(x)$ is the normalized dc bias current
density and $\Gamma$ is the normalized magnetic field. In
accordance with RSJ model \cite{lik},\cite{barone} one takes the
loss parameter $\alpha=\frac{\omega_{p}}{\omega_{c}}$, where
$\omega_p=\sqrt{2eI_c/\hbar C}$, $\omega_{c}=2eI_cR_{N}/\hbar$,
$C$ is the capacitance, $R_N$ is the normal state resistance
($R_N=V/I_{qp}$, $V$ being voltage and $I_{qp}$ -- the
quasiparticle component of the current), $I_c$ is the critical
current, $\eta(x)=J(x)/J_c$ ($I=\int\limits_0^l J(x)dx$,
$I_c=\int\limits_0^l J_c(x)dx$, $I$ is the bias current), $l$ is
the dimensional length of the junction, $L=l/\lambda_J$.

As it has been mentioned in the introduction, the flux-flow
regime is characterized by the fulfilling of the following
conditions: $\eta_0/\alpha \gg 1$ and (or) $\Gamma\gg 1$, where
$\eta_0=\int_0^L\eta(x)dx/L$ is the dimensionless total dc
bias current.

Instead of linear mode theory and
perturbative analysis around rotating background
\cite{kulik},\cite{cgssv98}, \cite{ss99},
one can use more general Poincare method \cite{lwp}:
obtain the solution as the
series with respect to the naturally arising in the flux-flow
regime small parameter
$\epsilon=\left(\frac{\alpha}{\eta_0}\right)^2\ll 1$.
Let us change variables in Eq. (\ref{PSGE}),
$\tau=\frac{\eta_0}{\alpha}t$, $z=\frac{\eta_0}{\alpha}x$:
\begin{equation}
\frac{\partial^2\phi}{\partial\tau^2}+
\beta \frac{\partial\phi}{\partial\tau}-
\frac{\partial^2\phi}{\partial z^2}=
\frac{\beta}{\eta_0}\eta(z)-\epsilon\sin (\phi),
\label{SEI}
\end{equation}
where $\beta=\alpha^2/\eta_0$.

The steady-state solution of this equation may be
found in the form:
$\phi(\tau)=\phi_0(\tau)+\epsilon\phi_1(\tau)+
\epsilon^2\phi_2(\tau)+\dots$
($|\phi_0(\tau)|\gg\epsilon|\phi_1(\tau)|\gg
\epsilon^2|\phi_2(\tau)|\gg\dots$).
Substituting this into Eq. (\ref{SEI}) one can find the zero
order equation:
\begin{equation}
\frac{\partial^2\phi_0}{\partial\tau^2}+
\beta\frac{\partial\phi_0}{\partial\tau}-
\frac{\partial^2\phi_0}{\partial z^2}=\frac{\beta}{\eta_0}\eta(z).
\label{SE0}
\end{equation}
It is easy to see, that the steady-state solution of this equation
is: $\phi_0(\tau)=\tau+\gamma z -g(z)= \frac{\eta_0}{\alpha}t+\Gamma
x -g(x)$, $\gamma=\alpha\Gamma/\eta_0$, where:
\begin{equation}
g(x)=\sum_{m=1}^{\infty}\frac{2}{Lk_m^2}\int_0^L\eta(x)\cos
k_mxdx,
\end{equation}
$k_m=\pi m/L$, so in the $0$-order approximation the current-voltage
characteristic is given by the ohmic line:
$v(\eta_0)=d\bar{\phi}/dt=\Omega_J(\eta_0)=\eta_0/\alpha$
(due to Josephson relation the voltage is proportional to the
Josephson frequency $\Omega_J$, here $\bar{\phi}$ means averaging
in time). To get higher order equations let us decompose
$\sin(\phi_0(\tau,z)+\epsilon\phi_1(\tau,z)+
\epsilon^2\phi_2(\tau,z)+\dots)$ into Taylor expansion. From the
structure of the considered linear recurrent equations it is
clear that the steady-state solution $\phi_n(\tau,z)$ may be
presented in the form: $\phi_n(\tau,z)=\omega_n
\tau+\phi_{np}(\tau,z)$, where $\phi_{np}(\tau,z)$ is periodic
nongrowing component.

Collecting together all linearly growing components
$\omega_n\tau$ one can get: $\sin(\phi(\tau,z))=
\sin(\{\omega_0\tau+\epsilon\omega_1\tau+\epsilon^2\omega_2\tau
+\dots+\gamma z-g(z) \}+\epsilon\phi_{1p}(\tau)+
\epsilon^2\phi_{2p}(\tau)+\dots)$. Now one can linearize
$\sin(\phi)$ as: $\sin(\phi)\approx \sin(\omega_J\tau+\gamma
z-g(z))+\epsilon\phi_{1p}(\tau,z) \cos(\omega_J\tau+\gamma
z-g(z))+ \epsilon^2\phi_{2p}(\tau,z) \cos(\omega_J\tau+\gamma
z-g(z))+\dots$, where
$\omega_J=\omega_0+\epsilon\omega_1+\epsilon^2\omega_2+\dots$ is
the oscillation frequency ($\omega_0=1$), and $\omega_1$,
$\omega_2$,..., $\omega_n$,..., $\phi_{1p}(\tau,z)$,
$\phi_{2p}(\tau,z)$, ..., $\phi_{np}(\tau,z)$, ... are unknown
functions that is required to obtain. One can consider the
solution up to the 4-th order (in \cite{kulik}-\cite{ss99} the
solution up to the 2-nd order has been derived, but in
principle it can be done up to any order, all equations may be
solved recursively), and then the following equations for
$\phi_{1}(\tau,z)$ -- $\phi_{4}(\tau,z)$ may be written:
\begin{equation}
\frac{\partial^2\phi_1}{\partial\tau^2}+\beta
\frac{\partial\phi_1}{\partial\tau}-\frac{\partial^2\phi_1}{\partial z^2}=
-\sin(\omega_J\tau+\gamma z-g(z)),
\label{SE1}
\end{equation}
\begin{equation}
\frac{\partial^2\phi_2}{\partial\tau^2}
+\beta\frac{\partial\phi_2}{\partial\tau}
-\frac{\partial^2\phi_2}{\partial z^2}=
-\phi_{1p}(\tau)\cos(\omega_J\tau+\gamma z-g(z)),
\label{SE2}
\end{equation}
\begin{equation}
\frac{\partial^2\phi_3}{\partial\tau^2}
+\beta\frac{\partial\phi_3}{\partial\tau}
-\frac{\partial^2\phi_3}{\partial z^2}=
-\phi_{2p}(\tau)\cos(\omega_J\tau+\gamma z-g(z)),
\label{SE3}
\end{equation}
\begin{equation}
\frac{\partial^2\phi_4}{\partial\tau^2}
+\beta\frac{\partial\phi_4}{\partial\tau}
-\frac{\partial^2\phi_4}{\partial z^2}=
-\phi_{3p}(\tau)\cos(\omega_J\tau+\gamma z-g(z)).
\label{SE4}
\end{equation}

The boundary conditions (\ref{bc}) are taken into account
in the solution of the zero order equation and Eqs.
(\ref{SE1})-(\ref{SE4}) should be solved for zero boundary
conditions:
$\left.\frac{\partial\phi}{\partial x}\right|_{x=0,L}=0$.
The solutions of linear Eqs. (\ref{SE1})-(\ref{SE4}) may
easily be derived recursively, substituting the solution in the
form: $\phi_n(\tau,z)=\sum\limits_{m=0}^\infty C_{nm}(\tau)
\cos\overline{k}_{m}z$. The computer simulations of Eq.
(\ref{PSGE}) have been performed on the basis of an implicit
difference scheme (similar to one, presented in the Appendix
of \cite{Zhang}) with adaptively varying time step and
calculation time.

\section{The 4-th order approximation for the homogeneous case}

Since the peculiarities of the derivation of the 2-nd order
approximation (solution of equations (\ref{SE1}),(\ref{SE2})) for
homogeneous case have been considered in detail in \cite{lwp}, the
only solution given in original notations of Eqs.
(\ref{PSGE}),(\ref{bc}) is presented below. The current-voltage
characteristic $v(\eta_0)=\Omega_J(\eta_0)=d\bar{\phi}/dt$ may be
obtained in the form:
$\Omega_J(\eta_0)=\eta_0/\alpha+\Omega_2(\eta_0)+\Omega_4(\eta_0)$, where
$\Omega_2(\eta_0)$ and $\Omega_4(\eta_0)$ are the second and the fourth order
corrections of frequency (voltage), respectively. As it has been
demonstrated in \cite{lwp}, one has to solve transcendental
equations for $\Omega_2$ and $\Omega_4$ in order to derive the
voltage $\Omega_J$ as function of the bias current $\eta_0$.
However, in homogeneous case one can express the bias current
$\eta_0$ as function of $\Omega_J$:
$\eta_0(\Omega_J)=\alpha\left\{\Omega_J-\Omega_2(\Omega_J)
-\Omega_4(\Omega_J)-\dots\right\}$ that
gives an analytical expression for $\eta_0(\Omega_J)$ and allows
to avoid the solution of the transcendental equation.

For the homogeneous case the 2-nd order correction as
function of $\Omega_J$ may be presented in the form:
\begin{equation}
\Omega_2=-\sum\limits_{n=0}^{\infty}
\frac{2-\delta_{0,n}}{2}\left[\frac{\Omega_J
[I_{Sn}^2+I_{Cn}^2]}{(\alpha\Omega_J)^2+
[k_n^2-\Omega_J^2]^2}\right],
\label{w2h}
\end{equation}
\begin{equation}
I_{Sn}=\frac{\Gamma L (1-\cos(\Gamma L)\cos(\pi
n))}{(\Gamma L)^2-(\pi n)^2}, \,
I_{Cn}=\frac{\Gamma L \sin(\Gamma L)\cos(\pi n)}{(\Gamma
L)^2-(\pi n)^2}.
\label{Ih}
\end{equation}

Correspondingly, the 4-th order correction may be written as:
\begin{equation}
\Omega_4=-\frac{1}{2\alpha}\sum\limits_{i=0}^{\infty}
\left[A_{31i}I_{Ci}-B_{31i}I_{Si}\right],
\label{w4}
\end{equation}

\begin{eqnarray}
A_{31i}=\frac{2-\delta_{0,i}}{2}
\frac{\alpha\Omega_J\sum\limits_{m=0}^{\infty}
\left[A_{22m}J_{Smi}+B_{22m}J_{Cmi}\right]-
(k_{i}^2-\Omega_J^2)\sum\limits_{m=0}^{\infty}
\left[A_{22m}J_{Cmi}-B_{22m}J_{Smi}\right]}
{(\alpha\Omega_J)^2+(k_{i}^2-\Omega_J^2)^2}
\nonumber
\end{eqnarray}
\begin{eqnarray}
B_{31i}=-\frac{2-\delta_{0,i}}{2}
\frac{\alpha\Omega_J\sum\limits_{m=0}^{\infty}
\left[A_{22m}J_{Cmi}-B_{22m}J_{Smi}\right]+
(k_{i}^2-\Omega_J^2)\sum\limits_{m=0}^{\infty}
\left[A_{22m}J_{Smi}+B_{22m}J_{Cmi}\right]}
{(\alpha\Omega_J)^2+(k_{i}^2-\Omega_J^2)^2}
\nonumber
\end{eqnarray}

\begin{eqnarray}
A_{22m}=\frac{2-\delta_{0,m}}{2}
\frac{2\alpha\Omega_J\sum\limits_{n=0}^{\infty}
\left[B_{1n}J_{Cnm}-A_{1n}J_{Snm}\right]-
(k_{m}^2-4\Omega_J^2)\sum\limits_{n=0}^{\infty}
\left[A_{1n}J_{Cnm}+B_{1n}J_{Snm}\right]}
{(2\alpha\Omega_J)^2+(k_{m}^2-4\Omega_J^2)^2}
\nonumber
\end{eqnarray}
\begin{eqnarray}
B_{22m}=-\frac{2-\delta_{0,m}}{2}
\frac{2\alpha\Omega_J\sum\limits_{n=0}^{\infty}
\left[A_{1n}J_{Cnm}+B_{1n}J_{Snm}\right]+
(k_{m}^2-4\Omega_J^2)\sum\limits_{n=0}^{\infty}
\left[B_{1n}J_{Cnm}-A_{1n}J_{Snm}\right]}
{(2\alpha\Omega_J)^2+(k_{m}^2-4\Omega_J^2)^2}
\nonumber
\end{eqnarray}

\begin{eqnarray}
A_{1n}=(2-\delta_{0,n})\frac{\alpha\Omega_J I_{Cn}-
({k}_{n}^2-\Omega_J^2)I_{Sn}}
{(\alpha\Omega_J)^2+({k}_{n}^2-\Omega_J^2)^2},
\nonumber
\end{eqnarray}
\begin{eqnarray}
B_{1n}=-(2-\delta_{0,n})\frac{\alpha\Omega_J I_{Sn}+
({k}_{n}^2-\Omega_J^2)I_{Cn}}
{(\alpha\Omega_J)^2+({k}_{n}^2-\Omega_J^2)^2},
\nonumber
\end{eqnarray}

\begin{eqnarray}
J_{Cnm}=\frac{1}{2}\left\{\frac{\Gamma L \sin\Gamma L\cos\pi(m+n)}
{(\Gamma L)^2-[\pi(m+n)]^2}+\frac{\Gamma L \sin\Gamma L\cos\pi(m-n)}
{(\Gamma L)^2-[\pi(m-n)]^2}\right\},
\nonumber
\end{eqnarray}
\begin{eqnarray}
J_{Snm}=\frac{1}{2}\left\{\frac{\Gamma L(1-\cos\Gamma L\cos\pi(m+n))}
{(\Gamma L)^2-[\pi(m+n)]^2}+\frac{\Gamma L(1-\cos\Gamma L\cos\pi(m-n))}
{(\Gamma L)^2-[\pi(m-n)]^2}\right\}.
\nonumber
\end{eqnarray}

As one can guess, in order to obtain $J_{Cmi}$ and $J_{Smi}$,
one has to interchange indexes in $J_{Cnm}$ and $J_{Snm}$,
respectively; functions $I_{Sn}$ and $I_{Cn}$ are given by
(\ref{Ih}).

The current-voltage characteristic
$\eta_0(\Omega_J)=\alpha\left\{\Omega_J-\Omega_2(\Omega_J)
-\Omega_4(\Omega_J)\right\}$
is presented in Fig. 1 for $\alpha=0.1$, $L=5$, $\Gamma=3$.
It follows from the analysis that if the deviation from the ohmic
line is located in the area where $\eta_0/\alpha>2$,
it is usually enough to use the 2-nd order approximation only. It
is seen, that the account of the fourth order correction gives
more precise description of the height of Fiske steps. For
description of Fiske or Eck steps that occur below the
boundary $\eta_0/\alpha>2$ the 4-th order correction
becomes of importance and significantly improves the
approximation, see the inset of Fig. 1, where it is seen that the
4-th order approximation describes the step missed in the
2-nd order approximation (around $\Omega_J\approx 0.95$). It
should be noted that for larger magnetic fields the second order
approximation coincides well with the results of computer
simulation for rather small damping (e.g. for $\alpha=0.04-0.01$,
see Fig. 2 for $\alpha=0.04$, $\Gamma=5$, $L=5$), but it is not
easy to catch all Fiske steps in computer simulations in this
case.

\section{The 2-nd order approximation for the inhomogeneous case}

For inhomogeneous case, unfortunately, one can not avoid the
solution of the transcendental equation and for simplicity let us
consider the inhomogeneous case in the second order approximation
only. For the second order correction $\Omega_2$ of the
current-voltage characteristic in the inhomogeneous case one can
get the following transcendental equation:
\begin{equation}
\Omega_2=-\sum\limits_{n=0}^{\infty}
\frac{2-\delta_{0,n}}{2}\left[\frac{(\eta_0/\alpha+\Omega_2)
[I_{Sn}^2+I_{Cn}^2]}{((\eta_0+\alpha\Omega_2))^2+
[k_n^2-(\eta_0/\alpha+\Omega_2)^2]^2}\right],
\label{wi2}
\end{equation}
\begin{equation}
I_{Sn}=\frac{1}{L}\int_0^{L}
\sin(\Gamma x-g(x))\cos(k_{n} x) dx, \,
I_{Cn}=\frac{1}{L}\int_0^{L}
\cos(\Gamma x-g(x))\cos(k_{n} x) dx.
\label{Iih}
\end{equation}
It should be noted, that both in the homogeneous and
inhomogeneous cases the equation for $\Omega_2$ is
given by the same formula (\ref{wi2}) with the only difference
that in homogeneous case $g(x)=0$ in (\ref{Iih}) and functions
$I_{Sn}$ and $I_{Cn}$ may be evaluated analytically (\ref{Ih}).

As an example of spatial distribution of the bias current let
us consider the unit-step function (unbiased tail):
$\eta(x)=\eta_s\sigma(x-x_0)$, $\eta_s=\frac{\eta_0 L}{L-x_0}$.
In this case ($k_{n}=\pi n/L$):
\begin{equation}
g(x)=\sum\limits_{n=1}^{\infty}
\frac{2\eta_s}{\pi n k_n^2}\sin k_nx_0 \cos k_n x=
\frac{\eta_s}{6L}\left\{
\begin{array}{ll}
(L-x_0)(x_0(2L-x_0)-3x^2), & x<x_0, \\
x_0(2L^2-6xL+3x^2-x_0^2), & x>x_0 \\
\end{array}
\right.
\label{gx}
\end{equation}

The transcendental equation (\ref{wi2}) with $g(x)$ given by
(\ref{gx}) may easily be solved and the voltage-current
characteristic of FFO $\Omega_J(\eta_0)=\eta_0/\alpha+\Omega_2(\eta_0)$ may be
found. In Fig. 3 the results of computer simulation of Eq.
(\ref{PSGE}) are presented for $\alpha=0.3$, $L=5$, $\Gamma=5,-5$,
$x_0=2.5$ (long unbiased tail) and $x_0=0$ (no
unbiased tail). One can see, that in the case where the driving
is homogeneous ($x_0=0$) the current-voltage characteristic
is absolutely antisymmetric with respect to the bias current
and the sign of magnetic field $\Gamma$ does not play any role.
Contrary, in the case with unbiased tail ($x_0=2.5$) the
current-voltage characteristic becomes asymmetric:
if the bias is applied at the end, where vortices
exit the junction, the radiation is amplified, if the bias
is applied at the end, where vortices enter the junction,
the radiation is suppressed. The comparison between
analytically derived IVC on the basis of Eq.
(\ref{wi2}) and results of computer simulation is presented
in the inset of Fig. 3 for $\alpha=0.3$, $L=5$, $\Gamma=-5$,
$x_0=2.5$.

As it is seen from Fig. 4, not only the location of unbiased
tail, but also its length plays an important role: the amplitude
of the main harmonic has maximum as function of the length of
unbiased tail (for the considered parameters the maximum is
located around $x_0=2.5$; it follows from Eq. (\ref{SE2}) that the
second order correction of the IVC $\Omega_2$ is proportional to
the amplitude of the first harmonic $\phi_1(x,t)$).
In the inset of Fig. 4 the results of computer simulation of IVC
minus ohmic part is given for different values of $x_0$ and it is
also seen that the main resonance is higher for $x_0=2.5$.
It should be noted, however, that an optimal length of unbiased
tail will be different for different parameters of the long
Josephson junction (such as length, damping and magnetic field)
and special investigations should be performed for optimization
of designs of practical FFOs.

It is clearly seen from Fig. 4 and the inset of Fig. 3, that the
second order approximation of IVC in the inhomogeneous case gives
rather good coincidence with the results of computer simulation,
but the discrepancy increases with the increase of the length of
unbiased tail, which is due to the fact that the deviation of IVC
from ohmic line occurs for larger values of $\alpha/\eta_0$ and the
small parameter $\epsilon=(\alpha/\eta_0)^2$ becomes larger
leading to worse applicability of the approximation.

Let us perform the analysis of voltage-field characteristic of
FFO to study the dependence of $R_d^{H}=dV/dH$ (or
$r_d^{\Gamma}=dv/d\Gamma$ in dimensionless notation) on the
length of unbiased tail. From the plots of $\Omega_J(\Gamma)$,
presented in Fig. 5,
we do not see so significant dependence of $R_d^{H}$
from the length of the unbiased tail as it was for $R_d$.
For smaller bias current the curve $\Omega_J(\Gamma)$ consists
from several branches with positive derivative.
However, for larger currents $\eta_0\ge 1$ it is intriguing to see
that there is a range of parameters, where the derivative
($R_d^{H}$) takes negative values. As it has been demonstrated in
\cite{lwp}, if fluctuations of bias current $\eta_F(t)$ and
magnetic field $\Gamma_F(t)$ are correlated (such that
$\Gamma_F(t)=\sigma\eta_F(t)$, $\sigma$ is a numeric
coefficient), the linewidth $\Delta f$ of FFO is proportional to
$(R_d+\sigma R_d^{H})^2$ (where $R_d=dV/dI$,
$R_d^{H}=dV/dH$) and if $R_d<\sigma R_d^{H}$ one can see the
plato on the plot of $\Delta f(R_d)$ (see \cite{kosh2}), that is,
with further decrease of $R_d$ the linewidth does not change.
It is obvious that if $R_d>0$ and $\sigma R_d^{H}<0$ (meaning
that the noises of bias current and magnetic field are
anticorrelated) the linewidth may be significantly decreased.
Certainly, in real situations even if $|R_d|=|\sigma R_d^{H}|$
(or, alternatively, if $R_d$ and $\sigma R_d^{H}$ are both
positive, but very small), the linewidth will never be zero since other
noise sources (e.g., technical fluctuations) that were not taken
into account in \cite{lwp} would become of importance and
the saturation of $\Delta f(R_d,R_d^{H})$ should be observed.
It follows from estimations that FFO operating in
the considered range of parameters (small length and large
damping) will have rather small power that may be not enough for
practical applications. Nevertheless, since there is still a lack of
understanding about nature of magnetic field fluctuations, it
could be interesting to experimentally study this case for
$\sigma R_d^{H}<0$ to check if bias current and magnetic field
fluctuations are correlated or not. In the case of negative
$\sigma R_d^{H}$ this question may easily be resolved: if the
noises are correlated the linewidth is proportional to
$(R_d+\sigma R_d^{H})^2$ and may be significantly reduced if
$|R_d|\approx|\sigma R_d^{H}|$; if the noises are
uncorrelated, the linewidth is proportional to $R_d^2+(\sigma
R_d^{H})^2$ and the sign of $\sigma R_d^{H}$ will not
affect the linewidth. It could be interesting to study the
possibility to achieve stable generation regimes for
$\sigma R_d^{H}<0$ for practical FFOs, but as
preliminary results demonstrate the curves of $\Omega_J(\Gamma)$
are similar to that presented in Fig. 5 for $\eta_0=0.5$, i.e.
they have positive derivatives. However, the optimization of
practical FFO designs with respect to working parameters is out
of scope of the paper and will be presented elsewhere.

\section{Conclusions}

In the present paper the approach for description of the dynamics
of dense soliton chain in long Josephson junctions for the case
of large bias current and magnetic field is outlined.
The approach allows to perform the detailed analytical study
of flux flow oscillators with spatially inhomogeneous driving
and to find an optimal profile of the bias current distribution
in order to maximize the output power and to minimize the
linewidth. Practically important example of FFO with unbiased
tail has been considered and the existence of an optimal length
of the unbiased tail has been demonstrated. The possibility to
significantly reduce the linewidth by the design of FFOs with
anticorrelated bias current and magnetic field fluctuations has
been discussed.

\section{Acknowledgments}

The author wishes to thank V. P. Koshelets, V. V. Kurin, M. Yu.
Levitchev, J. Mygind and I. A. Shereshevsky for helpful
discussions. The work has been supported by the Russian
Foundation for Basic Research (Project N~99-02-17544, Project
N~00-02-16528 and Project N~00-15-96620) and by INTAS (Project
N~02-0367 and Project N~02-0450).

\newpage

\begin{figure}[th]
  \centering
\includegraphics[width=12cm,height=10cm]{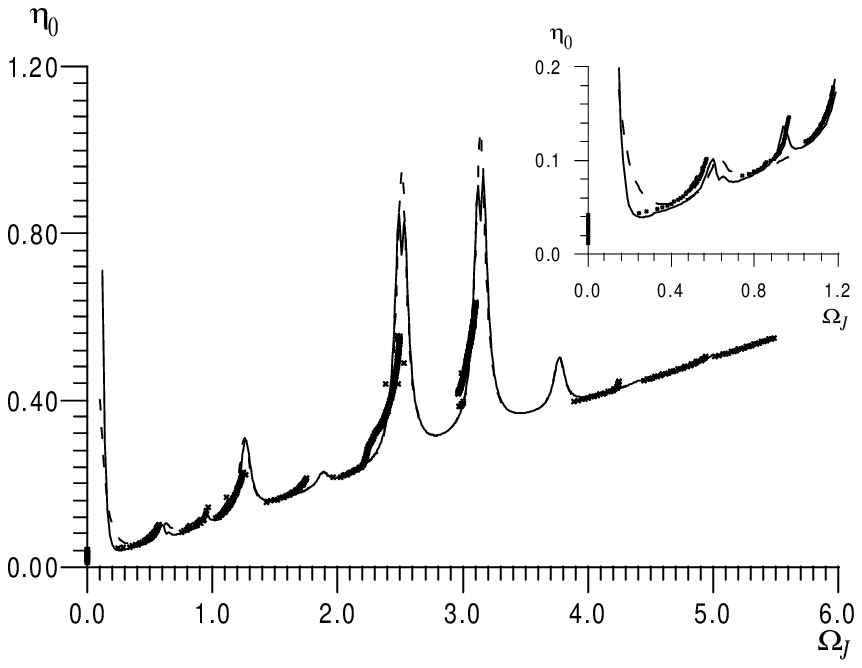}
\caption{Current-voltage characteristic of FFO
with homogeneous driving. Numerical solution
of the sine-Gordon equation is presented by crosses,
the fourth order approximation is given by solid
line and the second order approximation is given by
dashed line for $\alpha=0.1$, $L=5$, $\Gamma=3$.
Inset: Enlargement of the IVC for $\Omega_J<1.2$,
the step (around $\Omega_J\approx 0.95$) missed in the second
order is reproduced in the fourth order approximation
(dimensionless units).}
  \label{f1}
\end{figure}

\newpage

\begin{figure}[th]
  \centering
\includegraphics[width=12cm,height=10cm]{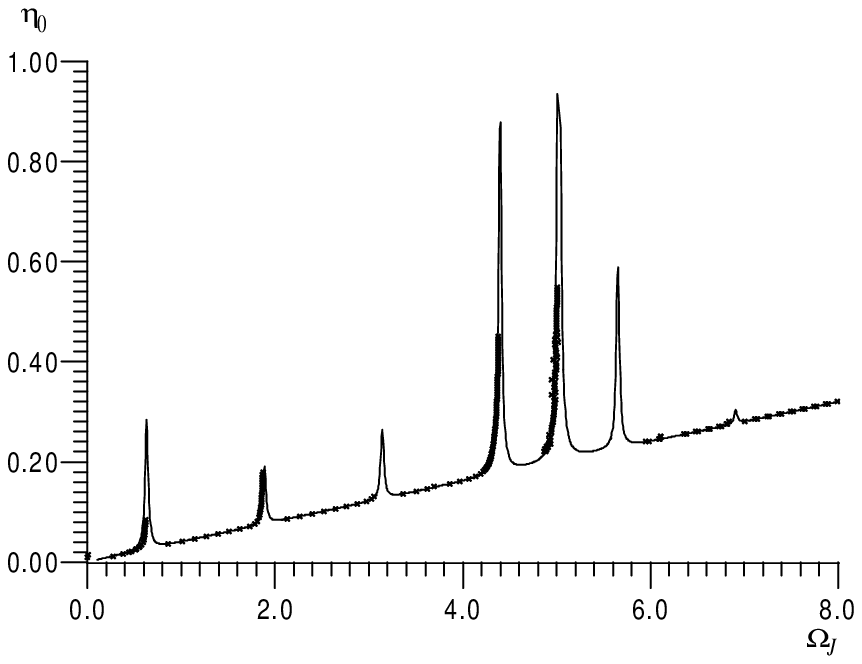}
\caption{Current-voltage characteristic of FFO with homogeneous driving
for $\alpha=0.04$, $L=5$, $\Gamma=5$; crosses - computer
simulations, solid line - theory (dimensionless units).}
  \label{f2}
\end{figure}

\newpage

\begin{figure}[th]
  \centering
\includegraphics[width=12cm,height=10cm]{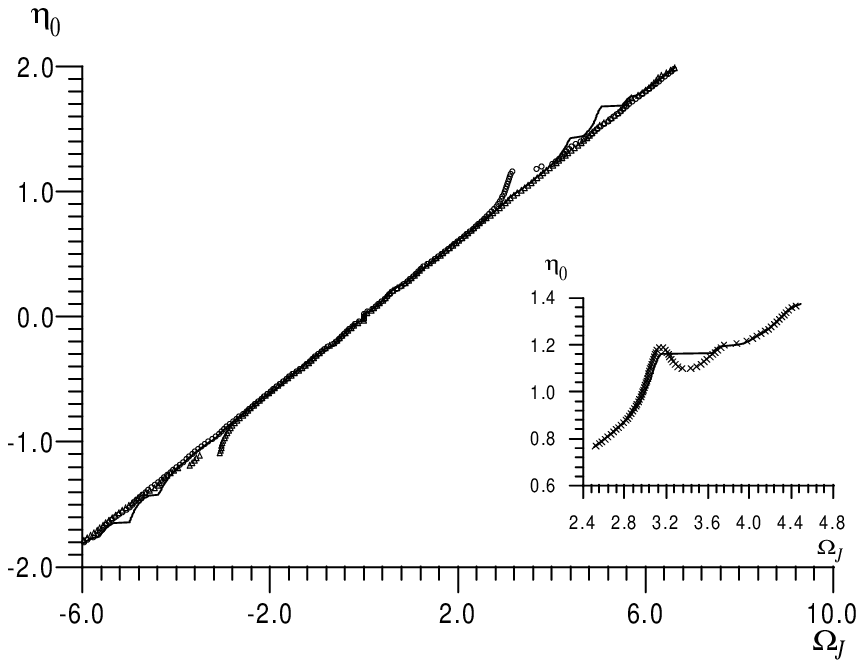}
\caption{Current-voltage characteristic of FFO with unbiased tail,
$\alpha=0.3$, $L=5$; for $x_0=0$, $\Gamma=5$ - solid line;
for $x_0=2.5$, $\Gamma=5$ - triangles; for $x_0=2.5$,
$\Gamma=-5$ - circles (see explanation in the text).
Inset: IVC for $\alpha=0.3$, $L=5$, $x_0=2.5$,
$\Gamma=-5$; solid line - computer simulations,
crosses - second order approximation (dimensionless units).}
  \label{f3}
\end{figure}

\newpage

\begin{figure}[th]
  \centering
\includegraphics[width=12cm,height=10cm]{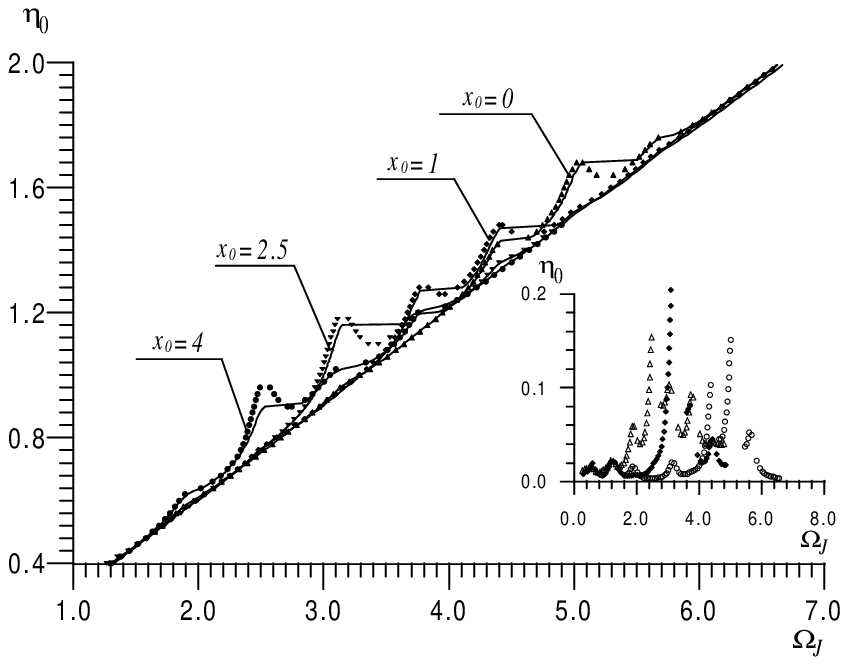}
\caption{Current-voltage characteristic of FFO with unbiased tail.
$\alpha=0.3$, $L=5$, $\Gamma=-5$, bias is from $x_0$ to $L$;
solid lines - computer simulation, dots - theory. It is seen,
that there is an optimal length of the bias tail both from
the point of view of $R_d$ and power of the main harmonic.
Inset: IVC minus ohmic line for $\alpha=0.3$, $L=5$, $\Gamma=-5$
and $x_0=0$ - open circles, $x_0=2.5$ - filled diamonds and
$x_0=4$ - open triangles (dimensionless units).}
  \label{f4}
\end{figure}

\newpage

\begin{figure}[th]
  \centering
\includegraphics[width=12cm,height=10cm]{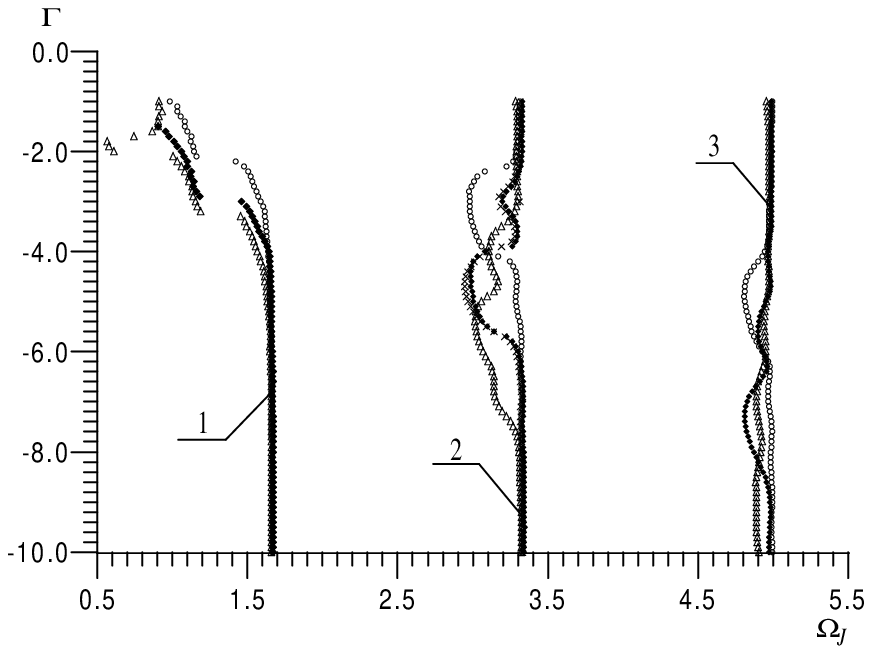}
\caption{Voltage as function of magnetic field $\Gamma$ for $\alpha=0.3$,
$L=5$ and different values of bias current: 1 - $\eta_0=0.5$, 2 -
$\eta_0=1$, 3 - $\eta_0=1.5$; open circles - $x_0=0$, filled
diamonds - $x_0=2.5$, open triangles - $x_0=4$; crosses - theory
for $\eta_0=1$, $x_0=2.5$ (dimensionless units).}
  \label{f5}
\end{figure}

\end{document}